\documentclass{elsart}
\usepackage{psfig}
\begin{document}
\runauthor{Cicero, Caesar and Vergil}
\begin{frontmatter}
\title{X-ray Fluctuations from the Slim Disk
}
\author[Osaka]{Mitsuru Takeuchi}

\address[Osaka]{Astronomical Institute, Osaka-kyoiku University, 
Kashiwara, 582-8582, Japan}
\begin{abstract}
The responses of perturbations added into the optically thick,
advection-dominated accretion disk (ADAD),
what we call the slim disk (SD),
are investigated
through numerical simulations.
Although it is proposed that the SD is thermally stable,
I find that a perturbation added into the disk is not rapidly damped
and moves through the disk in its free-fall time.
After the perturbation moves, the global structure of 
the disk does not vary very much.
These facts may account for
the substantial variability of the X-ray luminosities
of stellar super-luminal jet sources (SLJSs) and Narrow-Line
Seyfert~1s (NLS1s).
\end{abstract}
\end{frontmatter}

\section{Introduction}
Recent X-ray observations report
that not only stellar black hole candidates (SBHCs)
in their low states (=faint sources), 
but also NLS1s and
stellar SLJSs 
(=bright sources) exhibit
X-ray fluctuations (variability). 
The fluctuations of bright sources are made in 
the optically thick ADAD,
what we call SD (=bright disk).
However, the time evolution of the SD
has not been investigated so far. 
The numerical simulation by Manmoto et~al. (1996)
is well known as a time-evolution calculation of 
the optically thin ADAD (=faint disk).
The disturbance added into the optically thin
ADAD falls into the central star.
The disk luminosity increases when the disturbance falls,
and the light curve of this process
is in good agreement with the X-ray shot configuration
of the SBHC Cyg X-1 in its low state.

I investigate how the luminosities of disturbed SDs vary.
I add a similar disturbance as Manmoto et~al. (1996)
into the SD, and as a result,
a similar light curve to that of the optically thin
ADAD is obtained.

The basic
equations and numerical procedures are described in \S 2.
The resultant
time evolution and discussion will be presented in \S 3.


\section{Basic Equations}
I calculate the evolution of a one-dimensional axisymmetric disk.
The basic equations are the same as those of
Manmoto et~al. (1996),
and are those
of mass conservation, momentum conservation, angular momentum conservation,
and energy flow: 

\begin{equation}
\label{mass}
   {\partial\over\partial t}(r\Sigma)
 + {\partial\over\partial r}(r\Sigma v_{\rm r}) = 0,
\end{equation}
\begin{equation}
\label{r-mom}
   {\partial\over\partial t}(r\Sigma v_{\rm r})
\!\!+\!\!{\partial\over\partial r}(r\Sigma v_{\rm r}^2) 
=-r{\partial W\over\partial r}
\!\!+\!\! r^2\Sigma (\Omega^2 - \Omega_{\rm K}^2)
 - W{d\ln\Omega_{\rm K}\over d\ln r},
\end{equation}
\begin{equation}
\label{ang-mom}
   {\partial\over\partial t}(r^2\Sigma v_{\varphi})
 + {\partial\over\partial r}(r^2\Sigma v_{\rm r} v_{\varphi})
=-{\partial\over\partial r}(r^2 \alpha W),
\end{equation}
and
\begin{equation}
\label{energy}
   {\partial\over\partial t}(r\Sigma e)
\!\!+\!\!{\partial\over\partial r}(r\Sigma ev_{\rm r})
 = -{\partial\over\partial r}(rW v_{\rm r})
\!\!-\!\!{\partial\over\partial r}(r\alpha W v_{\varphi}) - rF   
\end{equation}
where $\Sigma(\equiv \int\rho dz)$ is the surface density,
$W(\equiv \int p dz)$ is the vertically integrated pressure,
and $e$ is the internal energy of the accreting gas.
$\Omega (=v_{\varphi}/r)$ and $\Omega_{\rm K}
[=(GM/r)^{1/2}/(r-r_{\rm S})$]
are the angular frequency of the gas flow
and the Keplerian angular frequency
in the pseudo-Newtonian potential (Paczy\'nski \& Witta 1980),
respectively, where
$M$ is the mass of the central black hole
and $r_{\rm S}$ is the Schwarzschild radius.
I set the viscosity parameter to be $\alpha =0.1$.
%

\begin{table*}
\caption{Parameter sets of Models A--C.}
\label{t1}
\begin{center}
\end{center}
\begin{center}
\begin{tabular}{ccr}
\hline
Model  & $r_0$ & $k$   \\
\hline
\hline
A     & 20   & $-0.3$               \\
B      & 20    & $-0.1$          \\
C     & 50    & $-0.3$         \\
\hline
\end{tabular}
\\
\end{center}
\end{table*}

To evaluate the radiative cooling rate, $F$, 
I consider black body radiation:
\begin{equation}
\label{flux}
   F = \frac{8acT^4}{3\tau_{\rm R}/2 +\sqrt{3}},
\end{equation}
where $a$ is the radiation constant, $T$ is the gas temperature,
and $\tau_{\rm R}$ is the Rosseland optical depth.

I obtain the steady state solution of 
equations \ref{mass}--\ref{energy}
and perform the time-evolution calculation
using the steady state solution as the initial state.
As for the outer boundary condition,
I set all quantities to be those of 
the standard disk by Shakura \& Sunyaev (1973).
The inner boundary is set at $r_{\rm in}=2.7r_{\rm S}$,
where a free boundary is adopted.
I add a mass of 
$\dot M \triangle t$ 
into the disk through the outer
boundary at every time step ($\triangle t$).
I set $\dot M=100L_{\rm Edd}/c^2$
where $L_{\rm Edd}$ is the Eddington luminosity.

I add a perturbation (disturbance) to the initial state of the disk 
as
\begin{equation}
\frac{\delta \rho}{\rho}
         = k \exp\left[-\left(\frac{r-r_0}{\lambda/2}\right)^2\right].
\end{equation}
Throughout these calculations, I assign
the wavelength of the perturbation, $\lambda=20r_{\rm S}$.
I set two parameters, the radius of the center of the perturbation, $r_0$,
and the ratio between
$\delta \rho$ and $\rho$
at $r=r_0$, $k$.
These parameters are varied, with
three calculations being performed.
The parameter sets of the three calculations
are listed in Table~1.

\begin{figure}
\centerline{\psfig{figure=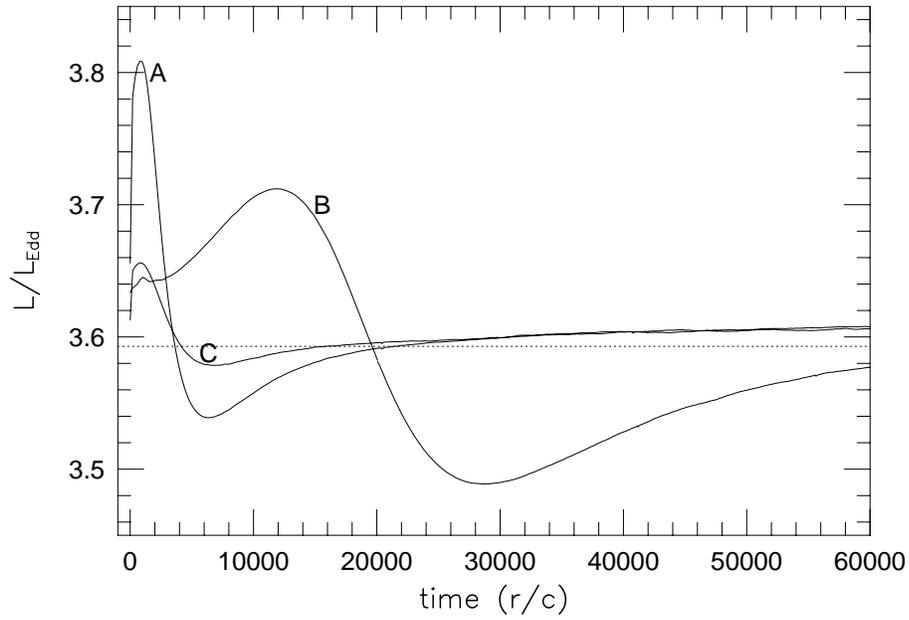,width=10cm,angle=90}}
\caption{Light curves of Models A--C.
The solid lines represent the light curves, and the
dotted line represents the luminosity in the steady state.
}
\label{f1}
\end{figure}

\section{Results and Summary}
Fig. 1 presents
the light curves of Models A--C.
All models show shot-like light curves.

\begin{figure*}
\centerline{\psfig{figure=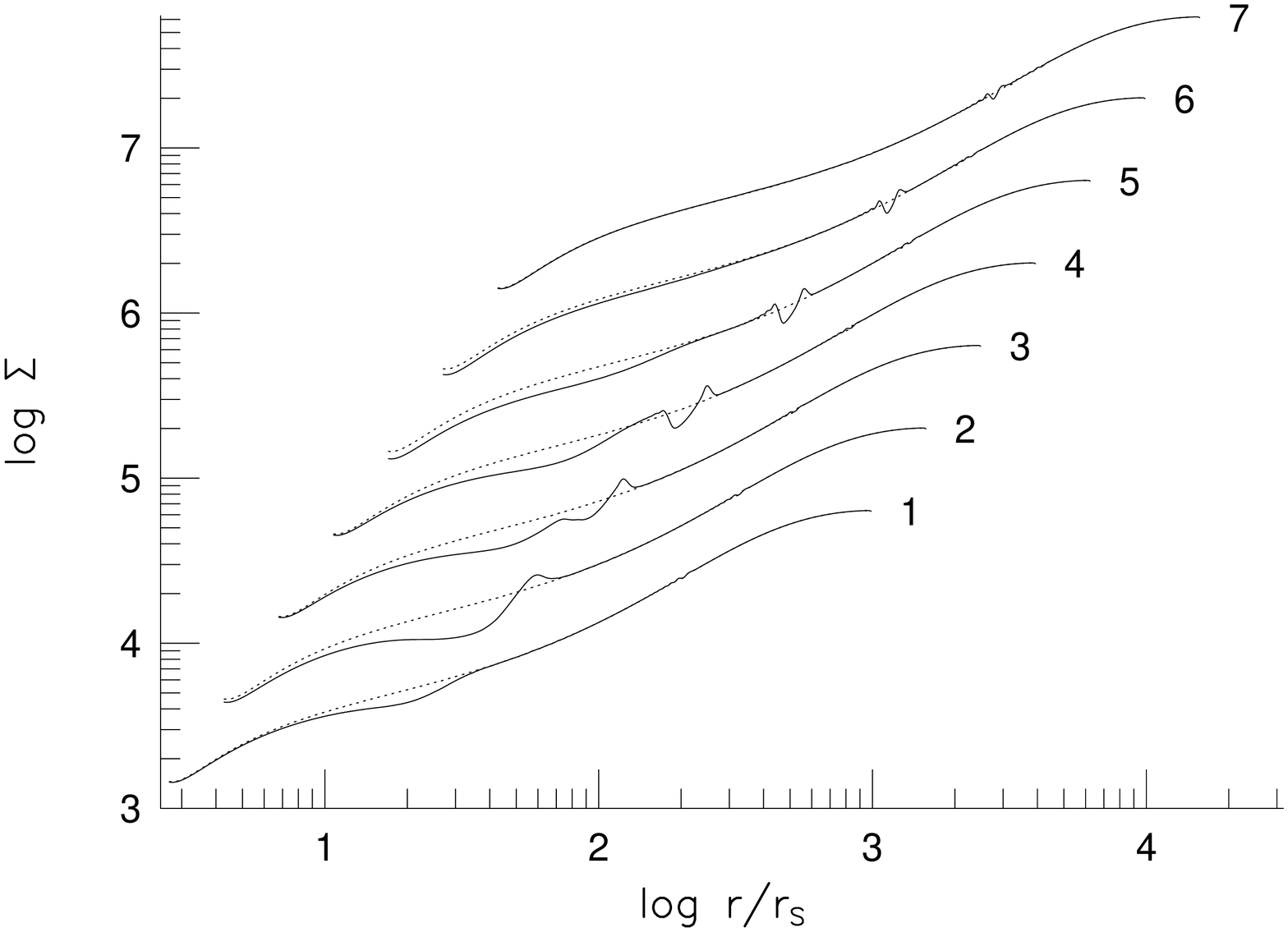,width=12cm,angle=90}}
\caption{
The time evolution of the disk surface density, $\Sigma$.
The dotted and solid lines represent
the initial state and the later evolution,
respectively.
The solid line labeled~1 means the initial configuration
of the disturbance.
Elapsed times for the labels 2--7 are
$t = 397,~785,~1352,~2894,~5997$ and $14288 r_{\rm S}/c$, 
respectively.
Note that both axes represent the relative values.
}
\label{f2}
\end{figure*}

Fig. 2 plots how the surface density,
$\Sigma$, varies after the perturbation is added for Model~A.
The dotted and solid lines represent
the initial state and the later evolution, respectively.
The perturbation 
does not change
its configuration very much, 
and it does not rapidly decay.
The time for the perturbation to propagate corresponds to
the free-fall time.
After the perturbation has damped, the disk structure 
is not globally changed.


The response of SDs to local
disturbances has been examined by one-dimensional numerical simulations.
It is generally believed that SDs are thermally stable. 
I, however, find that disturbances added into the 
accretion flow do not damp rapidly and decay
with roughly the free-fall time.
After the disturbance has damped, 
the global disk structure of the disk is not greatly modified.
This can account for the persistent X-ray emission with substantial
variations observed in NLS1s and 
SLJSs.
When a perturbation is made in the SD,
it decays
and exhibits one X-ray shot.
Since the structure of the SD does not globally vary much
after a perturbation propagates,
the next perturbation produced in the disk can also
exhibit an X-ray shot.
Repeating such processes continuously
can make the substantial variability seen in the X-ray
luminosities of SLJSs and NLS1s.

\end{document}